\documentclass[aps, pre, twocolumn, a4paper, superscriptaddress, floatfix,longbibliography]{revtex4-1}
\usepackage{graphicx}
\usepackage{amsmath}
\usepackage{placeins}
\usepackage{color}
\usepackage{listings}
\usepackage{array}
\newcommand{\er}{Erd\H{o}s-R\'{e}nyi~}
\newcommand{\kk}{\overline k }

\newcommand{\lcol}{\mathcal{L}_{\rm color}}

\newcommand{\beq}{\begin{equation}}
\newcommand{\eeq}{\end{equation}}
\newcommand{\beqa}{\begin{eqnarray}}
\newcommand{\eeqa}{\end{eqnarray}}

\begin{document}

\title{Color-avoiding percolation}

\author{Sebastian M. Krause}
\affiliation{Theoretical Physics Division, Rudjer Bo\v{s}kovi\'{c} Institute, Zagreb, Croatia}
\affiliation{Faculty of Physics, University of Duisburg-Essen, Duisburg, Germany}
\author{Michael M. Danziger}
\affiliation{Department of Physics, Bar Ilan University, Ramat Gan, Israel}
\author{Vinko Zlati\'{c}}
\affiliation{Theoretical Physics Division, Rudjer Bo\v{s}kovi\'{c} Institute, Zagreb, Croatia}
\affiliation{CNR Institute of Complex Systems UdR  Dep.~of Physics, University of Rome Sapienza, Rome, Italy}

\begin{abstract}
Many real world networks have groups of similar nodes which are vulnerable to the same failure or adversary. Nodes can be colored in such a way that colors encode the shared vulnerabilities. 
Using multiple paths to avoid these vulnerabilities can greatly improve network robustness. Color-avoiding percolation provides a theoretical framework for analyzing this scenario, focusing on the maximal set of nodes which can be connected via multiple color-avoiding paths.
In this paper we extend the basic theory of color-avoiding percolation that was published in [Krause~et.~al.,~Phys.~Rev.~X~6~(2016)~041022]. We explicitly account for the fact that the same particular link can be part of different paths avoiding different colors. This fact was previously accounted for with a heuristic approximation. We compare this approximation with a new, more exact theory and show that the new theory is substantially more accurate for many avoided colors. Further, we formulate our new theory with differentiated node functions, as senders/receivers or as transmitters. In both functions, nodes can be explicitly trusted or avoided. With only one avoided color we obtain standard percolation. With one by one avoiding additional colors, we can understand the critical behavior of color avoiding percolation. For heterogeneous color frequencies, we find that the colors with the largest frequencies control the critical threshold and exponent. Colors of small frequencies have only a minor influence on color avoiding connectivity, thus allowing for approximations.  
\end{abstract}
\maketitle

\section{Introduction}

In many real-world networks, the vulnerability of nodes to attack or failure is not uniform~\cite{helbing-nature2013}.
Instead, certain groups of nodes may share the same vulnerability, enabling a coloring of the nodes of the network in terms of their common vulnerabilities~\cite{krause2016hidden}.
For example, Internet routers run by the same entity or running the same software version may be subject to the same eavesdroppers \cite{ritchey2000using,choo2010cloud}, or multiple suppliers in a supply chain network may rely on the same critical resource \cite{SupplyChain1,SupplyChain2,SupplyChainRisk}.
In such cases, it may be desirable--or even essential--to find \textit{redundant} paths \cite{dolev-acm1993,stelling-cell2004,dorogovtsev-prl2006,goltsev-pre2006,carmi-pnas2007,newman-prl2008,radi-sensors2012,yeung2013physics} which \textit{avoid} every color.
This analysis can be translated into a new kind of percolation (about percolation see \cite{cohen-book2010,newman-book2010,morone-nature2015,peixoto2012evolution,PhysRevE.89.042811, zlatic2012networks, PhysRevE.94.032301}), called \textit{color-avoiding percolation}  which was first introduced in \cite{krause2016hidden}. Previously percolation theory on networks was among many other applications used to study robustness of complex systems \cite{cohen-prl2000,callaway2000network,gao-prl2011,danziger-collection2016}, epidemic spreading \cite{grassberger-mathbio1983,PhysRevE.66.016128,pastorsatorras-prl2001,serrano2006percolation}, opinion spreading \cite{goldenberg-physa2000}, and traffic \cite{wuellner2010resilience,li-pnas2015}.

In \cite{krause2016hidden}, we presented the theory of color-avoiding percolation based on probabilities that a randomly chosen node can (or cannot) communicate with a macroscopic fraction of other nodes over a particular link, avoiding a color $c$. 
These probabilities are dependent for different colors:
If a link is useful for communication avoiding one color $c$, it may be more likely useful for avoiding a second color $c'$ as well. 
These dependencies were treated with a heuristic approximation in \cite{krause2016hidden}. 
While this approximation works for all cases discussed there, its limits have not been discussed. Further, the dependencies have not been discussed in detail. 
Here we develop a theory treating dependencies explicitly. We show that while for few colors, the heuristic approximation of \cite{krause2016hidden} is suitable, when there are many colors it is not sufficient.
For every additional color to be avoided, dependencies of single link probabilities affect the reduction of color avoiding connectivity. In this way, we can understand the critical behavior of color avoiding percolation step by step, starting from standard percolation.

Promising generalizations of color avoiding percolation were introduced and applied to the autonomous systems level Internet: Heterogeneous color frequencies, and scenarios with trusted colors \cite{krause2016hidden}. Here we present a systematic discussion of these generalizations as well as new theoretical results as well.  We employ our theory with a flexible treatment of trust scenarios: Nodes of a certain color can be trusted or avoided as \textit{senders/receivers}, and they can be trusted or avoided for \textit{transmission}. This makes it possible to compare trust scenarios: Avoiding only one color (the color which is believed to be most likely to fail) for sending/receiving \emph{and} transmitting is equivalent to standard percolation. Avoiding more and more colors makes connectivity increasingly robust towards correlated failures, but possibly restricts the number of nodes which can participate.

We find a surprising and rich  critical behavior phenomenology. With heterogeneous color frequencies, we find that the avoided colors with the largest frequencies define the critical threshold and exponent. For colors with considerably smaller frequency, the difference in color avoiding percolation can be small, whether these colors are trusted or not. As a consequence, color avoiding connectivity can be increased by switching node colors such that dominant colors are reduced in frequency, or by increasing normal connectivity. The weak effect of small frequency colors on color avoiding percolation further allows us to introduce an approximation. This is important as, with the new, more detailed theory, the computational cost of calculating the largest color avoiding component increases exponentially with the number of avoided colors.

The structure of this paper is as follows: in Section \ref{sec:theory} we provide a full and precise treatment of color-avoiding percolation theory and we extend the theory to heterogeneous color frequencies, arbitrary sets of avoided transmitting node colors, and the case where sending node colors are also avoided.  
In Section \ref{sec:poisson} we focus on \er~ networks and develop a number of analytic results pertaining to the critical behavior of color-avoiding percolation including critical exponents and numerical validation of our analytic results.  
We also present analytic results for the phase transition which occurs when the color frequencies change. 
In Section \ref{sec:sender_receiver}, we study the case that each pair of sender and receiver nodes trust nodes of their respective colors, leading to a kind of inter-color adjacency matrix of color-avoiding connectivity.
In Section \ref{sec:approximations}, we describe different approximation approaches for calculating color-avoiding connectivity and explain the previously published approximation.

\section{Theory} \label{sec:theory}

\subsection{Color-avoiding percolation}

\begin{figure}[htb]
\begin{center}
    \begin{minipage}{0.49\columnwidth}\centering
    (a)\\    \includegraphics[trim=50 0 40 0,clip,width=1.0\textwidth]{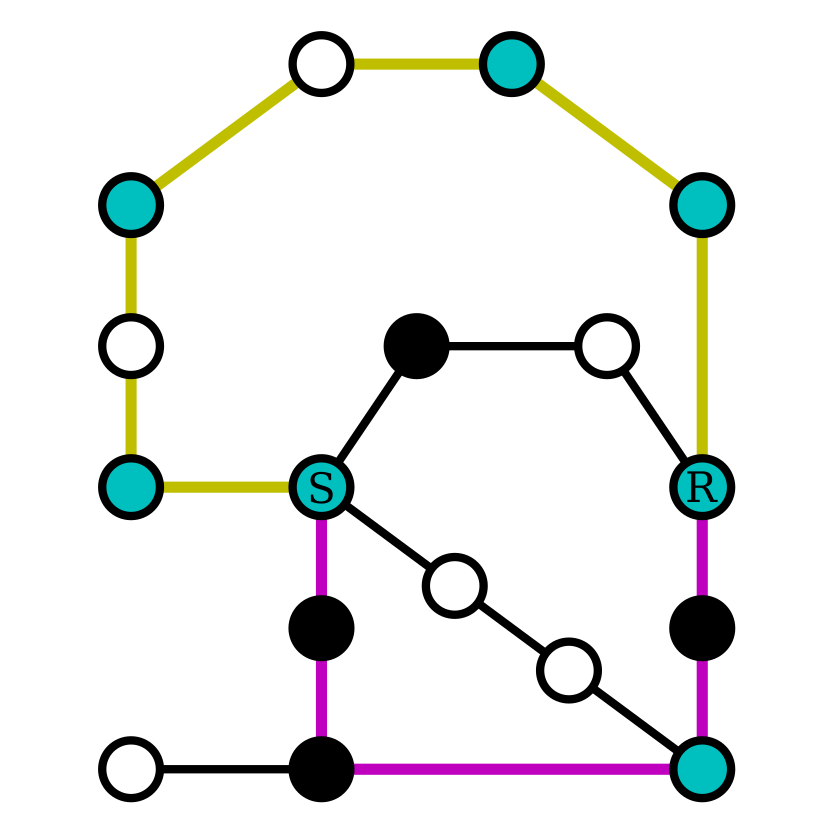}
    \end{minipage}
    \begin{minipage}{0.49\columnwidth}\centering
    (b)\\    \includegraphics[trim=40 0 50 0,clip,width=1.0\textwidth]{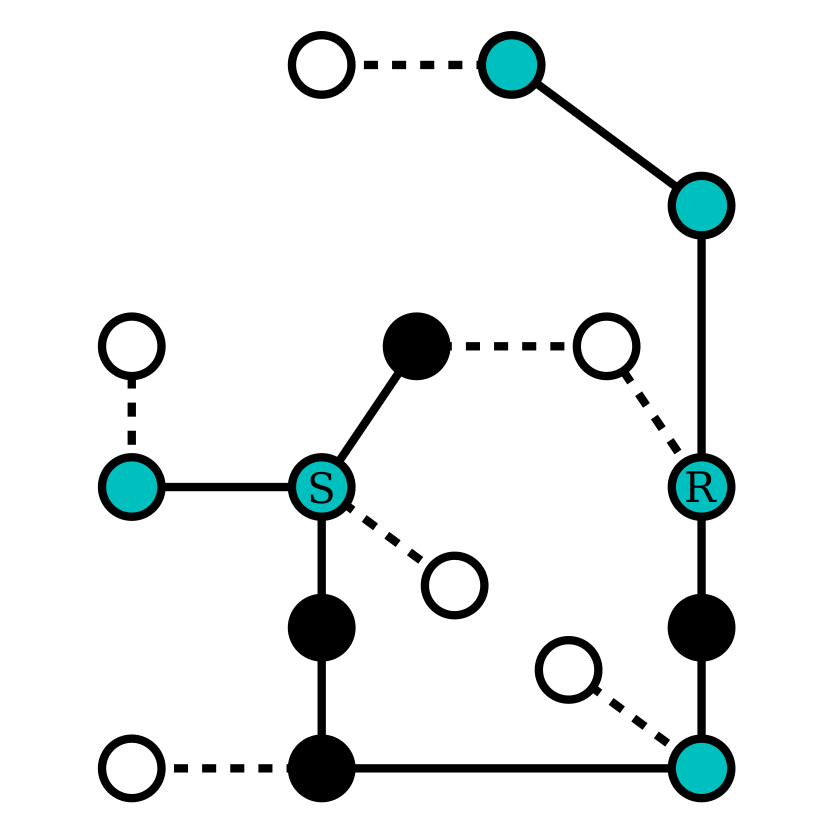}
    \end{minipage}
    \begin{minipage}{0.49\columnwidth}\centering
    (c)\\    \includegraphics[trim=50 0 40 0,clip,width=1.0\textwidth]{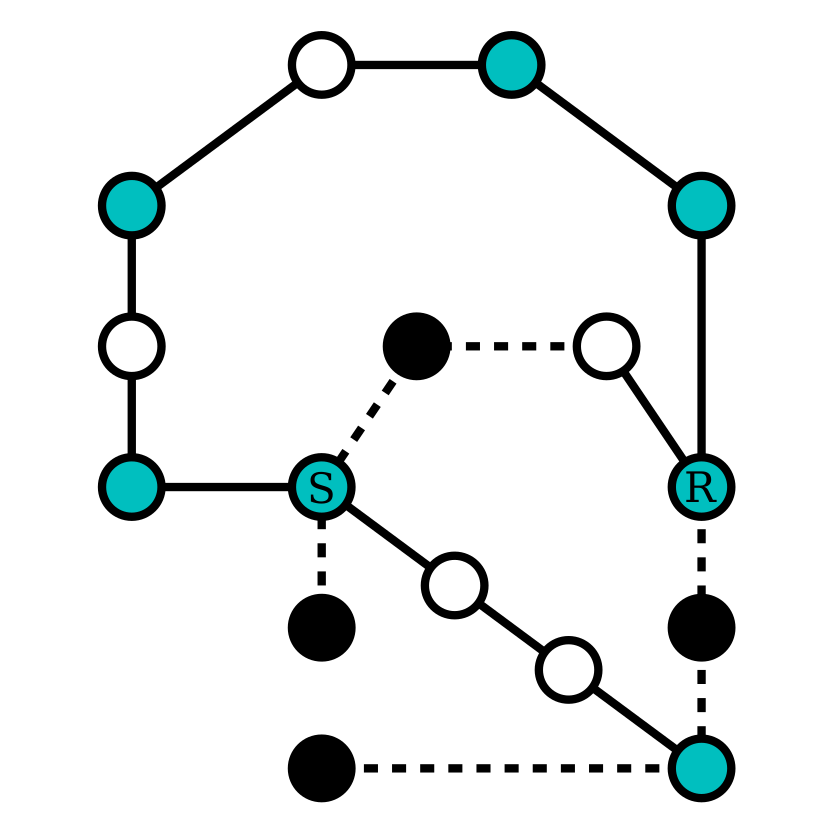}
    \end{minipage}
    \begin{minipage}{0.49\columnwidth}\centering
    (d)\\    \includegraphics[trim=40 0 50 0,clip,width=1.0\textwidth]{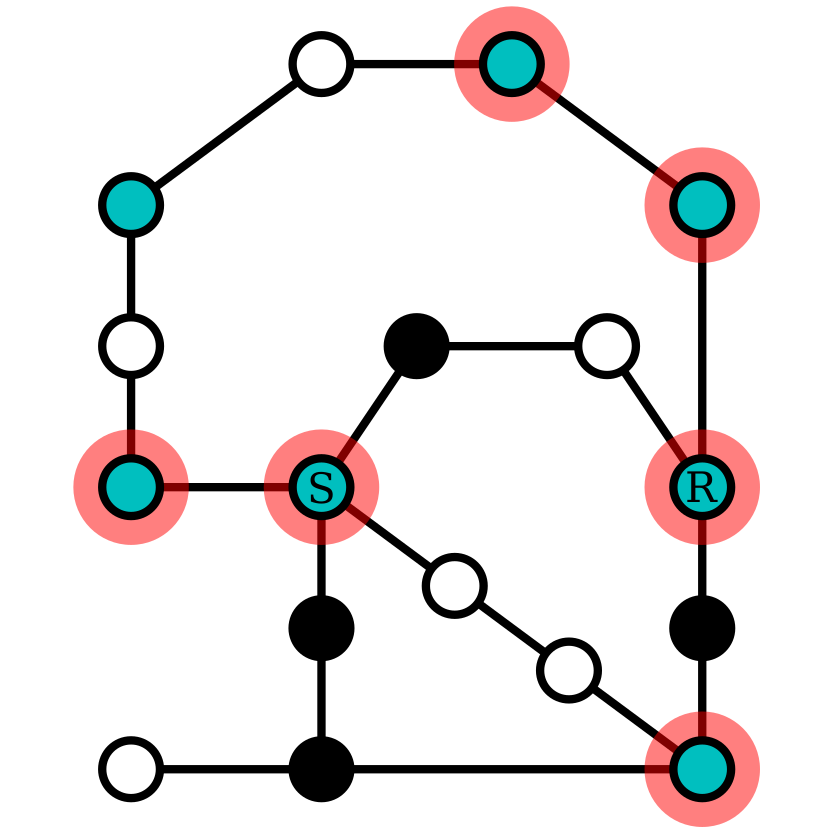}
    \end{minipage}
    \caption{{\bf Color avoiding percolation with redundant paths for avoiding white and black nodes.} {\bf (a)} Sender S and receiver R can communicate along the magenta path avoiding nodes of white color, or along the yellow path avoiding black nodes. {\bf (b)} The component $\mathcal{L}^+_{\overline{\rm white}}$, where every node pair can communicate over a path avoiding nodes of white color in between. Links to dangling nodes of white color are shown with dashed lines. {\bf (c)} Component $\mathcal{L}^+_{\overline{\rm black}}$. {\bf (d)} Nodes in  $\mathcal{L}_{\rm color}=\mathcal{C}_{\overline{\rm white}}\cap \mathcal{C}_{\overline{\rm black}} \cap \mathcal{L}^+_{\overline{\rm white}}\cap  \mathcal{L}^+_{\overline{\rm black}}$ are highlighted with red halos, where $\mathcal{C}_{\overline{c}}$ denotes the set of nodes with color other then $c$. Nodes in $\mathcal{L}_{\rm color}$ are pairwise color avoiding connected. 
    If black and white nodes would be avoided together, connectivity for blue nodes would break down.}
    \label{fig:CAP}
\end{center}
\end{figure}
We consider a network in which there are many different design problems or other vulnerabilities, and each problem affects nodes of a different color $c$.
This is illustrated in Fig. \ref{fig:CAP}a with three colors (cyan, white and black). 
Assuming there are $n_c$ bad colors (colors white and black in the figure) and $m_c$ good colors (cyan in the figure), we might naively opt to restrict connectivity to good colored nodes only.
However, this can render the entire system useless, if there are sufficiently many nodes with bad colors that their effective removal brings the network below the percolation threshold. 
For instance in Fig. \ref{fig:CAP}a, assuming that black and white nodes are bad, nodes S and R cannot communicate using only cyan nodes and are effectively disconnected. 
Instead of just removing all of the bad colors, we can look for redundant paths where each path avoids one of the bad colors. As we can see in Fig. \ref{fig:CAP}a, this can restore connectivity: Between nodes S and R there is a path avoiding black nodes (highlighted yellow) and a path avoiding white nodes (highlighted magenta).
Because no single of the bad colors is required for connectivity, we can restore connectivity while remaining robust to the failure of any single color.
Likewise, if the colors represent eavesdroppers and the message is split via secret sharing, and transmitted on multiple paths, no single eavesdropper can decode the whole message.

We define a pair of nodes as color-avoiding connected (CAC), if they have paths between them which avoid all of the vulnerable node colors, with each path avoiding a different node color.  
In the case that all of the colors are vulnerable, (as in \cite{krause2016hidden}), that means that we require as many paths as colors (though not necessarily distinct).
In Fig. \ref{fig:CAP}a we see that 
there exist paths between sender S and receiver R avoiding only white nodes or avoiding only black nodes, and they are thus CAC.
For color-avoiding percolation, we are interested in $\lcol$, the maximal set of nodes which are CAC, and the conditions under which $\lcol$ occupies a finite fraction of the total network.
In \cite{krause2016hidden}, the theory of color-avoiding percolation was limited to the case in which every node was colored and all of the colors were trusted as senders/receivers. For transmission, the focus was on the case where all colors were avoided. A generalization with trusted colors for transmission was introduced, but the theory was not discussed in more detail. Here we generalize that theory to the case where arbitrary sets of colors are avoided or trusted as transmitters or as senders/receivers.

Assuming $C$ disjunctive sets of node colors with frequencies $r_c$ such that $\sum_{c=1}^C r_c=1$,
we define the set of colors to be avoided as senders/receivers as $\mathcal{S}=\{s_1,\dots,s_{R}\}\subseteq \{1,\dots,C\}$, and the set of colors to be avoided as transmitters as $\mathcal{T}=\{t_1,\dots,t_{T}\}\subseteq \{1,\dots,C\}$. 
In Fig. \ref{fig:CAP}a we have colors (1,2,3)=(blue,white,black), and we want to avoid white and black nodes for sending and transmitting: $\mathcal{S}=\mathcal{T}=\{{\rm white},{\rm black}\}$.
If $\mathcal{S}=\mathcal{T}\neq\{1,\dots,C\}$, the node classes with colors in $\mathcal{T}$ are avoided for transmitting, and for sending and receiving as well. 
{ This can be used to represent the case, where certain node colors are universally mistrusted while other node colors are universally trusted.
For instance, if a node of color $c$ is attempting to communicate with a node of color $c'$, it may trust all nodes of its own color and the receiver color $c'$ but not trust any other colors as either sender/receivers or transmitters.
We return to this special case in detail in Sec.~\ref{sec:sender_receiver}}. 
For $\mathcal{S}=\emptyset$, nodes with problems are allowed to send or receive, while they are avoided for transmitting. This scenario is useful, if nodes of a class fail with a certain probability less than one. 
{ Or if the vulnerability of color $c$ does not impair the ability of the nodes to function as senders or receivers, but only impairs their transmitting abilities.}
In such a case, even though the sender itself has a vulnerable color, it still makes sense to avoid nodes of its own class on the path, in order to reduce the probability of disconnecting. 

We define $\mathcal{C}_{\overline{c}}$ as the set of all nodes with color \textit{other than} $c$, and $\mathcal{L}^+_{\overline{c}}$ as in \cite{krause2016hidden}: 
We remove all nodes with color $c$ and find the largest component in the remaining graph, $\mathcal{L}_{\bar c}$. 
We then define $\mathcal{L}^+_{\bar{c}}$ as the set of all nodes which have a direct neighbour in $\mathcal{L}_{\bar c}$. This trivially includes the whole $\mathcal{L}_{\bar c}$ and additionally includes nodes of color $c$ that are directly connected to $\mathcal{L}_{\bar c}$ ($\mathcal{L}_{\overline{\rm white}}^+$ in Fig. \ref{fig:CAP}b and $\mathcal{L}_{\overline{\rm black}}^+$ in Fig. \ref{fig:CAP}c).
These ``dangling'' $c$-colored nodes represent the nodes which can communicate via $\mathcal{L}_{\overline c}$, without requiring any $c$-colored nodes aside from themselves.

For the general case of avoiding a given set $\mathcal{S}$ of sender/receiver colors and a set $\mathcal{T}$ of transmitter colors, we find the color-avoiding giant component $\mathcal{L}_{\rm color}^{(\mathcal{S},\mathcal{T})}$ and its relative size $S_{\rm color}^{(\mathcal{S},\mathcal{T})}$ as
\begin{align}
\mathcal{L}_{\rm color}^{(\mathcal{S},\mathcal{T})} &= \mathcal{C}_{\overline{s_1}}\cap \dots \cap \mathcal{C}_{\overline{s_R}} \cap \mathcal{L}^+_{\overline{t_1}}\cap \dots \cap \mathcal{L}^+_{\overline{t_T}}\\
S_{\rm color}^{(\mathcal{S},\mathcal{T})} &= P(\mathcal{C}_{\overline{s_1}}\cap \dots \cap \mathcal{C}_{\overline{s_R}} \cap \mathcal{L}^+_{\overline{t_1}}\cap \dots \cap \mathcal{L}^+_{\overline{t_T}})\label{eq:S_color_prob}\\
 &= \left(1-\sum_{s\in\mathcal{S}} r_s\right) \underbrace{\left[1- P\left(\bigcup_{t\in\mathcal{T}}\neg\mathcal{L}^+_{\overline{t}}\right)\right]}_{S_{\rm color}^{\mathcal{T}}}.\label{eq:S_color_product}
\end{align}
With $\neg\mathcal{L}^+_{\overline{t}}$ we refer to the network of nodes \textit{not} belonging to $\mathcal{L}^+_{\overline{t}}$. Notice that negation turns the intersection into a union. The probability $S_{\rm color}^{(\mathcal{S},\mathcal{T})}$ describes the case where sending ($\mathcal{S}$) and transmitting ($\mathcal{T}$) nodes are avoided, and the probability $S_{\rm color}^{\mathcal{T}}$ describes the case where only transmitting nodes ($\mathcal{T}$) are avoided. 
In this manner, the giant color-avoiding component  $S_{\rm color}$ defined in \cite{krause2016hidden}, can be obtained by letting $\mathcal{T} = \{1,\dots,C\}$ and $\mathcal{S}=\emptyset$, where all colors are avoided for transmitting, and no color is avoided for sending and receiving. 

For Eq. \ref{eq:S_color_product}, color distribution and connectivity properties are assumed to be independent. The first term in this equation  describes a node property only, while the second term describes a property determined by connections and neighbors.  
We assume the colors to be distributed randomly, regardless of the network structure. 
This assumption has been verified with simulations, see for example Fig. \ref{fig:disjoint_C3}. 
In order to use the formalism of generating functions, which are not well suited to unions of probabilities, we use the inclusion-exclusion principle \cite{grinstead-book2012} to rewrite
\begin{align}
P\left(\bigcup_{t\in\mathcal{T}}\neg\mathcal{L}^+_{\overline{t}}\right) &= \sum_{\mathcal{Q}\subseteq\mathcal{T}} (-1)^{|\mathcal{Q}|-1} P\left(\bigcap_{q\in\mathcal{Q}}\neg\mathcal{L}^+_{\overline{q}}\right).\label{EQ:exc-inc}
\end{align}
$\mathcal{Q}$ takes on all possible subsets of $\mathcal{T}$, including $\mathcal{T}$ itself, but not the empty set. The term $|\mathcal{Q}|$ denotes the number of elements in the set $\mathcal{Q}$  which is a subset of the set of avoided colors.
In words, this equation uses the following rule: A sample is in the union of events, if it is in the first or second or third event etc., but double counting for pairwise intersections of events has to be subtracted. 
This procedure over-corrects intersections of triplets of events, which is then added back again and so on. 
We can now use single link probabilities and generating functions and rewrite 
\begin{align}
u_{\mathcal{Q}} &\equiv P_{1\rm link} \left(\bigcap_{q\in\mathcal{Q}}\neg\mathcal{L}^+_{\overline{q}}\right),\\
g_0(u_{\mathcal{Q}}) &= P\left(\bigcap_{q\in\mathcal{Q}}\neg\mathcal{L}^+_{\overline{q}}\right).\label{eq:first_with_generating}
\end{align}
$u_{\mathcal{Q}}$ is the probability, that a node belongs to $\neg \mathcal{L}^+_{\overline{q}}$ for all colors $q \in \mathcal{Q}$, after we destroyed all of its links except for one randomly chosen link. 
In other words, it gives the probability that a node is \textit{not} connected to \textit{any} of the components $\mathcal{L}_{\overline{q}}$ ($q \in \mathcal{Q}$) over one particular link. The set $\mathcal{Q}$ is meant as an index of $u_{\mathcal{Q}}$, so we will use $u_{\{1,2\}}$ for $u_{\mathcal{Q}}$ with $\mathcal{Q}=\{1,2\}$, or $u_{\mathcal{T}}$ for $\mathcal{Q}=\mathcal{T}$ etc. 
In Eq. \ref{eq:first_with_generating}, we assume random ensembles of networks with size going to infinity and the locally treelike approximation. The probability for all links to fail to connect a certain node to components $\mathcal{L}_{\overline{q}}$ can be found with the generating function of the degree distribution $g_0$.

Finally, we need equations defining probabilities $u_{\mathcal{Q}}$. We use self consistency equations: A link fails to connect a first node with probability $u_{\mathcal{Q}}$, if the other node along this link is not connected to any of $\mathcal{L}_{\bar{q}}$ for $q\in \mathcal{Q}$, over any of its other links. For the first color the usual equation holds:
\begin{align}
u_{\{c\}} &= r_c + (1-r_c) g_1(u_{\{c\}}),\label{Eq:self_cons_1}
\end{align}
where $g_1(z)=g_0'(z)/g_0'(1)$ is the generating function of excess degree \cite{newman-book2010}. For two colors $c\neq q$ the equation reads 
\begin{align}
u_{\{c,q\}} &= r_c g_1(u_{\{q\}})+r_q g_1(u_{\{c\}}) + \nonumber\\
&\qquad \qquad + (1-r_c-r_q) g_1(u_{\{c,q\}}).\label{Eq:self_cons_2}
\end{align}
In equation \ref{Eq:self_cons_2}, the first term represents the probability that the node reached through the link in question is of color $c$, and that at the same time it does not connect to the color avoiding component of color $q$. The second term represents the probability that the reached node is of color $q$, and that at the same time it does not connect to the color avoiding component of color $c$. The third term represents the probability that the reached node is neither of color $c$ nor $q$, but it also fails to connect to color avoiding components of colors $c$ and $q$ ( $L_{\overline{c}}$, $L_{\overline{q}}$). Clearly we have to plug in the result of the self consistent equations \ref{Eq:self_cons_1} into self consistent equation  \ref{Eq:self_cons_2}, to be able to compute the result. This has to be done numerically. 
In general, the equation for a joint probability that a node connects to \textit{none} of the components avoiding colors $\mathcal{Q}$ over a particular link reads 
\begin{align}
u_{\mathcal{Q}} &= \sum_{q\in\mathcal{Q}} r_q g_1(u_{\mathcal{Q}\setminus \{q\}}) + \left(1-\sum_{q\in \mathcal{Q}} r_q \right) g_1(u_{\mathcal{Q}})\label{eq:self_cons}
\end{align}
where $\mathcal{Q}\setminus \{q\}$ is defined as the set containing all colors included in $\mathcal{Q}$, except for the color $q$. 
For capturing the case of only one color $\mathcal{Q}=\{c\}$, we define $u_{\emptyset}=1$, which is a consistent definition, as it is a solution of \ref{eq:self_cons} itself ($u_{\emptyset}=g_1(u_{\emptyset})$). The only way to compute the joint probabilities for larger $|\mathcal{Q}|$ is to go step by step from joint probabilities for subsets of $\mathcal{Q'}\in\mathcal{Q}$ using equation \ref{eq:self_cons}. In this way, all the subsets of $\mathcal{T}$ have to be considered, including $\mathcal{T}$ itself. The results have to be plugged into the equation for $S_{\rm color}^{\mathcal{T}}$. Combining equation \ref{eq:S_color_product} and the following equations up to equation \ref{eq:first_with_generating} together, we find 
\begin{align}
S_{\rm color}^{\mathcal{T}} &= 1+\sum_{\mathcal{Q}\subseteq\mathcal{T}} (-1)^{|\mathcal{Q}|} g_0(u_{\mathcal{Q}}).\label{eq:S_color}
\end{align}
Further using equation \ref{eq:self_cons}, this gives us the necessary input for calculating the size of the giant color avoiding component, equation \ref{eq:S_color_product}. The comparison between this theory and the approximate theory developed in \cite{krause2016hidden} is laid out in section 3.D.

\subsection{Color-avoiding percolation as a generalization of standard percolation}

In standard site percolation, the question is whether a large part of a network stays connected after a random fraction of nodes is destroyed. 
To demonstrate how the color-avoiding percolation framework 
can be described as a generalization of standard percolation, we consider the following color-avoiding percolation problem which is equivalent to standard percolation.
We begin with a network composed of two node types: bad ($c=1$) and good ($c=2$).
We assume the bad fraction is known to have design problems (as an extra property beyond the network connections), and therefore should be avoided for connections. 
Each color occurs with frequency $r_c$ such that $r_1+r_2=1$. 
With $S$ we denote the fraction of nodes in the surviving giant component, or equivalently, the probability for a single node to belong to the surviving giant component. That means, all nodes of the bad color $c=1$ are excluded as senders/receivers and as transmitters. We obtain:
\begin{align}
S &= P(\mathcal{C}_{\overline{1}}\cap \mathcal{L}^+_{\overline{1}})\\
 &= P(\mathcal{C}_{\overline{1}}) P(\mathcal{L}^+_{\overline{1}})\label{EQ:2}\\
 &= P(\mathcal{C}_{\overline{1}}) [1- P(\neg \mathcal{L}^+_{\overline{1}})]\label{EQ:3}.
\end{align}
Here we have assumed that $\mathcal{C}_{\overline{1}}$ and $\mathcal{L}^+_{\overline{1}}$ are independent, as discussed for color avoiding percolation above. 
It holds $P(\mathcal{C}_{\overline{1}})=1-r_1$. Equations \ref{EQ:2} and \ref{EQ:3} are convenient for calculations using the formalism based on generating functions \cite{newman-book2010,newman-2001random}. Assuming random ensembles of networks with size going to infinity and the locally treelike approximation, we further obtain
\begin{align}
P(\neg \mathcal{L}^+_{\overline{1}}) &= g_0[P_{1{\rm link}}(\neg \mathcal{L}^+_{\overline{1}})],\\
 P_{1{\rm link}}(\neg \mathcal{L}^+_{\overline{1}}) &= r_1 + (1-r_1) g_1[P_{1{\rm link}}(\neg \mathcal{L}^+_{\overline{1}})].
\end{align}
$P_{1{\rm link}}(\neg \mathcal{L}^+_{\overline{1}})$ is the probability, that a node belongs to $\neg \mathcal{L}^+_{\overline{1}}$, after we destroyed all of its links except for one randomly chosen link. In other words, it gives the probability that a node is not connected to the giant component over one particular link. These equations are equivalent to equations that describe random attack on networks \cite{callaway2000network,cohen-prl2000}, as they should be.

\subsection{Notes on computation}

To summarize the framework of color avoiding percolation, we found that $2^{|\mathcal{T}|} -1$ transcendental equations \ref{eq:self_cons} have to be solved, to find $u_{\mathcal{Q}}$ for all subsets $\mathcal{Q}\subseteq\mathcal{T}$. Results are plugged into equation \ref{eq:S_color}. For illustrating the procedure, let us discuss an example with three total colors, of which two are avoided for sending and transmitting: $C=3$, $\mathcal{S}=\mathcal{T}=\{1,2\}$. 
\begin{align}
u_{\{1\}} &= r_1 + (1-r_1) g_1(u_{\{1\}}),\nonumber\\
u_{\{2\}} &= r_2 + (1-r_2) g_1(u_{\{2\}}),\nonumber\\
u_{\{1,2\}} &= r_1 g_1(u_{\{2\}}) + r_2 g_1(u_{\{1\}}) + r_3 g_1(u_{\{1,2\}}),\nonumber\\
S_{\rm color}^{(\mathcal{S},\mathcal{T})} &= r_3 \left[ 1-g_0(u_{\{1\}}) -g_0(u_{\{2\}}) +g_0(u_{\{1,2\}})\right].\label{eq:three_colors_disjoint}
\end{align}
For ten avoided colors, $1023$ subsets of $\mathcal{T}$ have to be considered, which is numerically still easy to do. 
If frequencies of all avoided colors are identical, the number of different transcendent equations reduces drastically to $|\mathcal{T}|$. However, for tens of avoided colors, the limited precision of numerical results for $u_{\mathcal{Q}}$ (especially due to the limited precision numerics for the generating function $g_1$) and the combination of many terms limits feasibility of the straight forward evaluation. 
In the general case of heterogeneous color frequencies, we observed that summing over the sets needed is an easy numerical task for our computer as long as $|\mathcal{T}|\leq 20$. 
For twenty avoided colors, about one million subsets have to be considered. 
We provide Python code in the appendix which works fast and precise enough for all examples discussed in the following (for 20 disjoint colors, it needs about a minute to calculate $S_{\rm color}^{\mathcal{T}}$ on one core of an 
Intel\textregistered\ Core\texttrademark\ i7-4770 CPU (3.40GHz)), even if there is still much room for optimization. Below we discuss an approximation for reducing the number of colors, if many color frequencies are small compared to the largest color frequency.

\subsection{Criticality}

The critical behavior of color avoiding percolation has a rich, multifaceted nature, depending on graph topology, number of avoided colors and color frequencies. 
We will discuss a number of phenomena using special cases below. However, the transition point can be understood within the general theory framework, by referencing standard percolation. 
Without loss of generality, we assume for the avoided colors $\mathcal{T}=\{1,2,\dots,T\}$, and the total frequency of color $c=1$,  $r_1$, is larger or equal compared to the frequencies of all other avoided colors. We have for $c\in\mathcal{T}$: $u_{\{c\}}=r_c + \left(1-r_c\right)g_1(u_{\{c\}})$. From standard percolation we know, that if color $c=1$ can be avoided ($u_{\{1\}}<1$), then this is true for all other avoided colors as well, as they have smaller frequencies. Accordingly, color $c=1$ is the first to have vanishing connectivity, $u_{\{1\}}=1$. In this latter case, we have $S_{\rm color}^{(\mathcal{S},\mathcal{T})}=0$, as all terms $g_1(u_{\mathcal{Q}})$ in equation \ref{eq:S_color} cancel out pairwise. 
 First this is true for $1-g_1(u_{\{1\}})$. 
Equation \ref{eq:self_cons} for $u_{\{1,2\}}$ reduces with $g_1(u_{\{1\}})=1$ to the defining equation for $u_{\{2\}}$, thus $g_1(u_{\{1,2\}})-g_1(u_{\{2\}})=0$. 
 This generalizes to $u_{\mathcal{Q}}=u_{\mathcal{Q}\cup\{1\}}$. 
Accordingly, the critical point is determined by the avoided color with the largest frequency, in the same way as for standard percolation. With the result of Cohen~\cite{cohen-prl2000} we find as a condition for non-vanishing connectivity (with expected degree $\bar{k}$)
\begin{align}
r_c < r_{\rm crit} &= 1-\frac{\bar{k}}{\left<k^2\right>-\bar{k}}. 
\end{align}

\subsection{Dual variables for simultaneous probabilities}

The probabilities $u_{\mathcal{Q}}$ are hard to interpret, as they describe a negative statement: The probability of a link to \textit{not} connect to several $\mathcal{L}_{\bar c}$ at the same time. This makes it confusing to discuss the meaning of dependencies, for example in the form $u_{\{1,2\}}\neq u_{\{1\}} u_{\{2\}}$.  Furthermore, it turns out that for larger sets $\mathcal{Q}$ with more than one element, it is hard to find good approximations for $u_{\mathcal{Q}}$. This complicates understanding the critical behavior. Therefore, let us define new positive variables, describing the probability that a link connects to all $\mathcal{L}_{\overline{q}}$ for $q\in Q$ at the same time:
\begin{align}
v_{\mathcal{Q}} &= P_{1\rm link} \left(\bigcap_{q\in\mathcal{Q}}\mathcal{L}^+_{\overline{q}}\right)\\
&= 1+\sum_{\mathcal{P}\subseteq \mathcal{Q}} (-1)^{|\mathcal{P}|} u_{\mathcal{P}},\label{eq:vu}\\
u_{\mathcal{Q}} &= 1+\sum_{\mathcal{P}\subseteq \mathcal{Q}} (-1)^{|\mathcal{P}|} v_{\mathcal{P}}.\label{eq:uv}
\end{align}
The conditions between $u_{\mathcal{Q}}$ and $v_{\mathcal{Q}}$ hold with the complementary event and the inclusion-exclusion principle. As the transformation from $u_{\mathcal{Q}}$ to $v_{\mathcal{Q}}$ can be reversed with the same transformation (it is an involution), and it preserves all information, variables $v_{\mathcal{Q}}$ are dual variables to $u_{\mathcal{Q}}$. 
If probabilities were independent, to connect to several $\mathcal{L}_{\overline{q}}$ for $q\in Q$ at the same time over the same link, it would hold $v_{\mathcal{Q}}=\prod_{q\in\mathcal{Q}} v_{\{q\}}$. As we will discuss below, conditional probabilities for further colors can both be suppressed and increased. 

\begin{figure*}[htb]
\begin{center}
    \begin{minipage}{0.295\textwidth}\centering
    (a)\\    \includegraphics[width=1.0\textwidth]{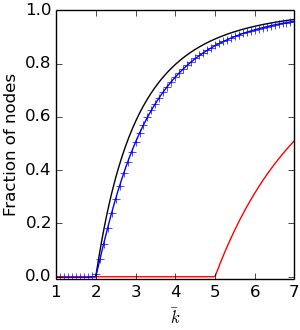}
    \end{minipage}
    \begin{minipage}{0.295\textwidth}\centering
    (b)\\    \includegraphics[width=1.0\textwidth]{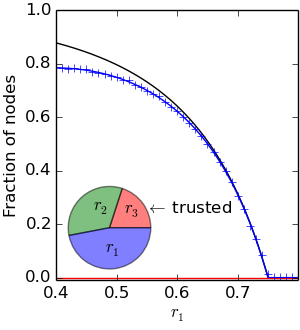}
    \end{minipage}
    \begin{minipage}{0.39\textwidth}\centering
    (c)\\    \includegraphics[width=1.0\textwidth]{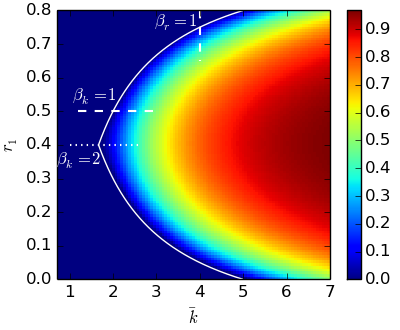}
    \end{minipage}
    \caption{{\bf Color avoiding connectivity depends on topology and color distribution.} Results for Poisson graphs, where the first two colors with frequencies $r_1+r2=0.8$ are avoided, and connectivity is measured as the fraction of nodes being CAC among the nodes of the trusted third color with $r_3=0.2$. 
    {\bf (a)} The black line shows the connectivity $S_{\rm color}^{\{1\}}$ avoiding the dominating color $c=1$ with frequency $r_1=0.5$. The blue line shows results $S_{\rm color}^{\{1,2\}}$, if first and second color are avoidable, where the second color has a smaller frequency $r_2=0.3$. The fraction of connected nodes reduces only slightly, even if many more nodes have to be avoided. Numerical results confirm the theory (blue crosses, averages over 100 networks of size $N=10^6$). 
    If nodes of colors one and two would be avoided at the same time, as shown with the red line, connectivity would reduce considerably. 
    {\bf (b)} For $\bar{k}=4$, color one cannot be avoided at all as long as $r_1>r_{\rm crit}=3/4$. Therefore, in a network with one dominating node class, it can be beneficial to replace colors of some nodes, thus reducing $r_1$ and increasing $r_2$ (as outlined in the inset). While standard connectivity stays suppressed (red line), color avoiding connectivity (blue line) increases almost as much as connectivity avoiding only the larger color (black line). 
    {\bf (c)} $S_{\rm color}^{\{1,2\}}$ is shown for varying topology and color frequencies. The solid white line indicates the critical manifold. Dashed white lines highlight the parameter manifolds of (a) and (b), and the phase transition along these lines has critical exponent $\beta_k=\beta_r=1$. The phase transition along the dotted white line is of different type, with exponent $\beta_k=2$ (see figure \ref{fig:crossover_C3}). This is connected to a sharp bend in the critical manifold. 
    }
    \label{fig:disjoint_C3}
\end{center}
\end{figure*}
It is useful to have equations for calculating $v_{\mathcal{Q}}$ directly, instead of first calculating $u_{\mathcal{Q}}$ and then transforming to $v_{\mathcal{Q}}$. Approximations are easier found for calculating $v_{\mathcal{Q}}$ directly. For $\mathcal{Q}\neq \{1,\dots,C\}$ it holds: 
\begin{align}
g_1(u_{\mathcal{Q}}) &= g_1\left(1+\sum_{\mathcal{P}\subseteq \mathcal{Q}} (-1)^{|\mathcal{P}|} v_{\mathcal{P}}\right)\\
&= 1+\sum_{\mathcal{P}\subseteq \mathcal{Q}} (-1)^{|\mathcal{P}|} \frac{v_{\mathcal{P}}}{1-\sum_{q\in \mathcal{P}}r_q}.\label{eq:self_cons_v}
\end{align}
In the first of these equations, only $u_{\mathcal{Q}}$ was plugged in using equation \ref{eq:uv}. Equation \ref{eq:self_cons_v} can be understood as follows: Starting with $\mathcal{Q}=\{c\}$, replace the left hand side of Eq. \ref{eq:self_cons} using Eq. \ref{eq:uv}, and solve for $g_1(u_{\{c\}})$. For increasing from $\mathcal{Q}\setminus{\{q\}}$ to $\mathcal{Q}$, replace the left hand side of Eq. \ref{eq:self_cons} using Eq. \ref{eq:uv}. Then plug in the lower order results already obtained to the right hand side of equation \ref{eq:self_cons}, and solve for $g_1(u_{\mathcal{Q}})$. Finally, it holds 
\begin{align}
   v_{\{1,\dots,C\}} &= 0.
\end{align}
This reflects the fact that a single link can never be used for avoiding all colors at the same time. The node reached over this link has a certain color, consequently this color cannot be avoided.

\section{Phase transition for Poisson graphs} \label{sec:poisson}

We start with equations \ref{eq:three_colors_disjoint}, where the first two out of three colors are avoided. In figure \ref{fig:disjoint_C3}, results are shown for Poisson graphs. Colors one and two are avoided for sending and for transmitting as well, and connectivity is calculated as the fraction of CAC nodes among the nodes of color three, $S_{\rm color}^{(\{1,2\},\{1,2\})}/r_3=S_{\rm color}^{\{1,2\}}$. For this figure, we fixed $r_1+r_2=0.8$ and $r_3=0.2$, and varied (a) connectivity with constant color frequencies ($r_1=0.5$ and $r_2=0.3$), (b) color frequencies of the avoided colors with fixed connectivity $\bar{k}=4$, and (c) both together. In (a) and (b), additional to the CAC results (blue lines) we show results similar to standard percolation, where only one color is avoided for transmission. 
With black lines we show results for avoiding only the dominant color for transmission, $S_{\rm color}^{(\{1,2\},\{1\})}/r_3=S_{\rm color}^{\{1\}}$. As above, the fraction among all nodes of color $c=3$ is shown. 
With red lines we show the situation where nodes of colors one and two are both together totally avoided for transmission, $S(r_1+r_2)/r_3$. Here we use the notation $S(\phi)$ for standard percolation, for a random fraction $\phi$ of destroyed / avoided nodes (fraction $1-\phi$ of surviving nodes respectively, compare equation \ref{EQ:3}). 
We see that CAC is almost as high as standard connectivity avoiding the dominant color, while avoiding the problematic nodes of colors one and two all together, reduces the connectivity remarkably. In (b), we further see that color avoiding connectivity increases only slowly, when color frequencies come close, $r_1\approx r_2$. This observation can help for finding the best cost-benefit trade-off of replacing nodes colors, if there is a dominant color in the system. With (c) we see that for color avoiding connectivity, both graph topology and color distribution are crucial. Therefore, a combination of improving topology and color heterogeneity can be most efficient for increasing color avoidability. 

To prepare for approximate results, let us switch to variables $v_{\mathcal{Q}}$. For Poisson graphs, we have $g_1(x)=g_0(x)=\exp[\bar{k}(z-1)]$ for the generating functions, with expected degree $\bar{k}$. Therefore, we have to solve equations \ref{eq:self_cons} for $g_1(u_{\mathcal{Q}})$, allowing to plug in $g_0(u_{\mathcal{Q}})=g_1(u_{\mathcal{Q}})$ directly into equation \ref{eq:S_color}. This is possible for $\mathcal{Q}\neq \{1,\dots,C\}$. Inserting results formulated with positive variables $v_{\mathcal{Q}}$ (Eq. \ref{eq:self_cons_v}) to equation \ref{eq:S_color}, we find for $\mathcal{T}\neq \{1,\dots,C\}$ that 
\begin{align}
S_{\rm color}^{(\mathcal{S},\mathcal{T})} &= \left(1-\sum_{s\in\mathcal{S}} r_s\right) \frac{v_{\mathcal{T}}}{1-\sum_{t\in\mathcal{T}} r_t}.\label{eq:S_color_v}
\end{align}
This is a surprisingly simple result, reducing to single link probabilities (the same holds for normal percolation on Poisson graphs, being captured with a single avoided color: $S=S_{\rm color}^{(\{1\},\{1\})}=1-u_{\{1\}}$). Finally, for $\mathcal{Q}=\{1,\dots,C\}$ we find the condition on $v_{\{1,\dots,C\}}=0$, instead of a replacement of $g_1$, which is not available here.

\subsection{Critical behavior}

For Poisson graphs, the conditions for critical parameters read 
\begin{align}
r_c < r_{\rm crit} &= \frac{\bar{k}-1}{\bar{k}} , \quad c\in\mathcal{T}
\end{align}
and solving for $\bar{k}$, we have 
\begin{align}
\bar{k}_{\rm crit} &= \frac{1}{1-\max_{c\in\mathcal{T}} r_c}.
\end{align}
Critical behavior according to varying topology or varying color frequencies is, as usual, described using the critical exponent $\beta$. For clarity we use two exponents: 
\begin{align}
S_{\rm color}^{(\mathcal{S},\mathcal{T})} &\propto \left(\bar{k}-\bar{k}_{\rm crit}\right)^{\beta_k},\\
S_{\rm color}^{(\mathcal{S},\mathcal{T})} &\propto \left(r_{\rm crit}-\max_{c\in\mathcal{T}} r_c\right)^{\beta_r}.
\end{align}
This can be dominated by the avoided color with the largest frequency, but in general the interplay of different colors can lead to new effects which are not present in standard percolation. 

The critical behavior of color avoiding percolation shows interesting features, as we see in Fig. \ref{fig:disjoint_C3}c. While the phase transition along the dashed white lines has critical exponent $\beta_k=\beta_r=1$, the situation is different, along the dotted white line, with $\beta_k=2$. The behavior of $\beta_k$ can be summarized as follows: The critical exponent $\beta_k$ is determined by the degeneracy of the maximal color frequency, i.e., the number of colors which are tied for most common,
\begin{align}
\beta_k &= n_{\rm deg} = \sum_{c\in \mathcal{T}} \delta_{r_c, \max_{t\in \mathcal{T}} r_t}.
\end{align}
Below we will confirm this result using an approximate scheme, explicitly presenting results for up to three avoided colors. 
In \cite{krause2016hidden} it was reported that $\beta_k = C$, under the assumption that all $C$ colors are avoided for transmission, and they all have the same frequency. The more detailed analysis here confirms these results and shows the correct extension to cases where the colors have different frequencies. 

\begin{figure}[htb]
\begin{center}
    \includegraphics[width=1.0\columnwidth]{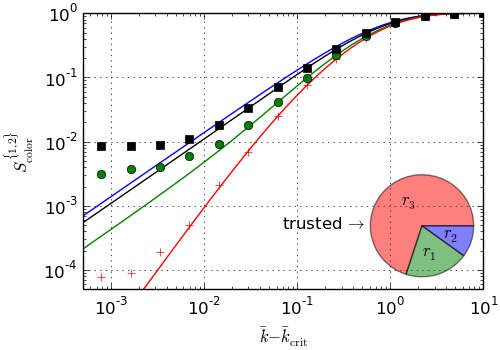}
    \caption{{\bf Crossover between critical behavior with $\beta_k=1$ (black line) and $\beta_k=2$ (red line).} We fix the frequency of the dominating avoided color $r_1=0.2$, thus fixing $\bar{k}_{\rm crit}=1.25$ on Poisson graphs. Results are shown for $r_2=r_1$ (red line), $r_2=r_1-0.03$ (green line), $r_2=r_1/2$ (black line) and $r_2=0$ (blue line, identical to standard percolation avoiding first color). For the third color, trusted for sending and transmitting, we have $r_3=1-r_1-r_2$. Symbols of according colors are averages over 100 networks of size $N=10^7$. }
    \label{fig:crossover_C3}
\end{center}
\end{figure}
The increased critical exponent $\beta_k=2$ along the dotted line in Fig. \ref{fig:disjoint_C3}c is connected to a sharp bend in the critical manifold, shown with a straight white line. It also is connected to the fact, that color connectivity is most suppressed compared to only avoiding the dominant color, if $r_1=r_2$ (see (b)). Between both kinds of phase transition, with $\beta_k=1$ and $\beta_k=2$, there has to be some kind of crossover. This is important to understand in order to determine the behavior of finite size networks with $r_1 \approx r_2$. 
Before we turn to an analytic examination of $S_{\rm color}$ in the crossover regime, let us discuss the phenomenology with figure \ref{fig:crossover_C3}. We use logarithmic scaling on both axis, thus power law dependencies show as straight lines. 
We set $r_1=0.2$ for the avoided color with the largest frequency, in order to keep $\bar{k}_{\rm crit}$ fixed. Results are shown with straight lines for the cases $r_2=r_1$ (red line), $r_2=r_1-0.03$ (green line), $r_2=r_1/2$ (black line) and $r_2=0$ (blue line). Numerical results are in good agreement with the theory, only limited by finite size effects for small size of the largest CAC component.  If the frequency of the second color is far below the frequency of the first color (black line), we see that color connectivity is overall reduced compared to one avoided color (blue line), but the critical behavior is not affected. If the frequency of the second color is close to that of the first color (green line), the behavior is close to this with identical color frequencies for large connectivity (red line). Therefore, the increased critical exponent $\beta_k=2$ is of practical relevance in finite size graphs, if color frequencies are close to each other. Closer to the critical point, the deviation of color frequencies results in a critical exponent of $\beta_k=1$. 

For small deviations of color frequencies $r_1\approx r_2$ (and $r_1\geq r_2)$, we can analyze the critical behavior analytically. We start with one avoided color. It is well known, that for Poisson graphs the critical exponent of standard percolation is $\beta_k=1$. We reproduce this result using the alternative formulation Eq. \ref{eq:self_cons_v} with an expansion of the generating function for small $v_{\{c\}}$ ($c=1,2$). $v_{\mathcal{Q}}$ replaces $u_{\mathcal{Q}}$ as defined in Eq. \ref{eq:vu}. We have 
\begin{align}
g_1(1-v_{\{c\}}) &= 1-\frac{v_{\{c\}}}{1-r_c}\\
&\approx 1 -\bar{k} v_{\{c\}} + (\bar{k}v_{\{c\}})^2/2\\
v_{\{c\}} &\approx \frac{1}{\bar{k}^2}\left(\bar{k}-\frac{1}{1-r_c}\right)\equiv \frac{2}{\bar{k}^2}\left(\bar{k}-\bar{k}_c\right).\label{eq:v1_app}
\end{align}
We define $\bar{k}_c=1/(1-r_c)$. For $\bar{k}<\bar{k}_1=\bar{k}_{\rm crit}$ we have $v_{\{1\}}=0$. With Eq. \ref{eq:S_color_v} we have $S_{\rm color}^{(\{1\},\{1\})}=v_{\{1\}}$, and therefore we found $\beta_k=1$.
How does the critical behavior generalize to two colors, $v_{\{1,2\}}$? Still, color $c=1$ is the color with higher frequency, and the critical behavior is connected to small $v_{\{1\}}$, increasing from zero. As we are interested in color frequencies with small deviation, we assume small $v_{\{2\}}$ as well. We develop the self consistency equation \ref{eq:self_cons_v} accordingly, 
\begin{align}
g_1(u_{\{1,2\}})&\approx 1-\bar{k} v_{\{1\}}+ \frac{(\bar{k}v_{\{1\}})^2}{2}-\bar{k} v_{\{2\}} + \frac{(\bar{k}v_{\{2\}})^2}{2} \nonumber\\
&\qquad + \bar{k}^2 v_{\{1\}} v_{\{2\}} +\bar{k}v_{\{1,2\}}\label{eq:v12_second_order}\\
&= 1-\frac{v_{\{1\}}}{1-r_1}-\frac{v_{\{2\}}}{1-r_2}+\frac{v_{\{1,2\}}}{1-r_1-r_2},\\
\bar{k}^2 v_{\{1\}} v_{\{2\}} &\approx v_{\{1,2\}} \left(\frac{1}{1-r_1-r_2}-\bar{k}\right),\label{eq:v_12_app}\\
v_{\{1,2\}} &\propto  (\bar{k}-\bar{k}_1)(\bar{k}-\bar{k}_2)\label{eq:v12_app}\\
 &= (\bar{k}-\bar{k}_{\rm crit}) [(\bar{k}-\bar{k}_{\rm crit})+(\bar{k}_1-\bar{k}_2)].
\end{align}
Most of the terms canceled out with approximate results for $v_{\{1\}}$ and $v_{\{2\}}$ from above. The largest correction in Eq. \ref{eq:v_12_app} is of order $v_{\{1\}} v_{\{2\}}^2$, therefore the approximation requires $v_{\{2\}}\ll1$. With $\bar{k}_1\approx \bar{k}_2$ and $S_{\rm color}^{(\mathcal{T},\mathcal{T})}=v_{\mathcal{T}}$, we find that color connectivity scales with $(\bar{k}-\bar{k}_{\rm crit})^2$, as long as $(\bar{k}-\bar{k}_{\rm crit}) \gg (\bar{k}_1-\bar{k}_2)$. For $r_1=r_2$, we thus find $\beta_k=2$. For $r_1>r_2$, the critical exponent $\beta_k=1$ is only visible for $(\bar{k}-\bar{k}_{\rm crit})\ll (\bar{k}_1-\bar{k}_2)$, what can be dominated by finite size effects, if color frequencies are close to each other.

\begin{figure}[htb]
\begin{center}
    \includegraphics[width=1.0\columnwidth]{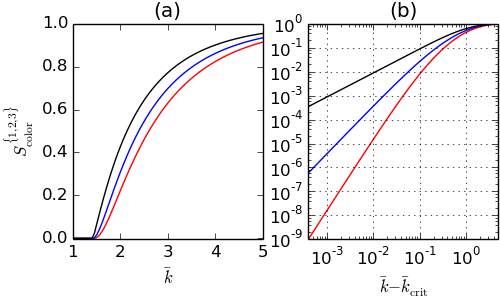}
    \caption{{\bf Influence of the dominating colors.} Results for Poisson graphs. The first three colors are avoided. 
    {\bf (a)} Compared to a single dominating color (black line, $r_1=0.3$, $r_2=r_3=0.12$, trusted fourth color with $r_4=0.46$), degenerated highest frequencies reduce the color avoiding connectivity stronger, and change the critical behavior (blue line: $r_1=r_2=0.3$, $r_3=0.12$, $r_4=0.28$; red line: $r_1=r_2=r_3=0.3$, $r_4=0.1$). The results are repeated in {\bf (b)} with logarithmic scaling. We see that the critical exponent $\beta_k$ is identical to the degeneration of the highest color frequency.}
    \label{fig:C4}
\end{center}
\end{figure}

With three avoided colors, we can analyze how the critical behavior generalizes to more colors. Figure \ref{fig:C4} shows results for disjoint colors on Poisson graphs, where three colors are avoided. We fix the frequency of the first color to $r_1=0.3$. Again, $S_{\rm color}^{\{1,2,3\}}$ can be seen as the fraction of nodes in the largest CAC component, among all nodes of color four. Compared to the black line, showing results where the other frequencies of avoided colors are smaller ($r_2=r_3=0.12$, $r_4=0.46$), degeneration of the highest avoided color frequency reduces color connectivity considerably. The results shown in (a) are repeated in (b) with logarithmic scaling, to demonstrate the critical power law behavior. The blue line shows results for double degenerate highest frequency $r_1=r_2=0.3$, and the exponent increases to $\beta_k=2$. We saw this exponent already for two avoided colors with degenerated frequencies. Now we see, that with a third avoided color with smaller frequency $r_3=0.12$, the exponent stays the same. The crossover as described for two avoided colors applies here as well, and the two-dimensional critical manifold in the parameter space $(r_1, r_2, r_3)$ has a sharp bend as well (results not shown). 
The red line shows results for triple degenerate highest color frequency of avoided colors $r_1=r_2=r_3$. This is connected with critical exponent $\beta_k=3$. This exponent can be extracted analytically, in the same way as it was done for two avoided colors. Using $v_{\{1\}}=v_{\{2\}}$ etc. for identical frequencies of avoided colors, we develop Eq. \ref{eq:v12_second_order} to third order (instead of second order), plug into an equation \ref{eq:self_cons_v} for $g_1(u_{\{1,2,3\}})$, develop to third order and solve the leading terms for $v_{\{1,2,3\}}$:
\begin{align}
v_{\{1,2,3\}} \left(\frac{1}{1-3 r_1}-\bar{k}\right) &\approx 3\bar{k}^2 v_{\{1\}} v_{\{1,2\}} + \frac{\bar{k}^3}{2} (v_{\{1\}})^3,\\
v_{\{1,2,3\}} &\propto (\bar{k}-\bar{k}_{\rm crit})^3,
\end{align}
where we have used the fact that $v_{\{1\}} \sim (\kk - \kk_{\rm crit})$ from Eq. \ref{eq:v1_app} and $v_{\{1,2\}} \sim (\kk - \kk_{\rm crit})^2$ from Eq. \ref{eq:v12_app}.
A similar procedure can be applied for higher numbers of avoided colors and heterogeneous color frequencies, which would require higher order expansions in the generating functions, and many terms of same order to be considered. The only case, in which this procedure is not applicable, is for avoiding all colors for transmission. It holds $v_{\{1,\dots,C\}}=0$, while $S_{\rm color}$ is finite. For this case, the approximation of \cite{krause2016hidden} can be useful, as discussed in the section about approximations.

\subsection{Dependencies among variables $v_{\mathcal{Q}}$ and critical exponents}

\begin{figure}[htb]
\begin{center}
    \includegraphics[width=1.0\columnwidth]{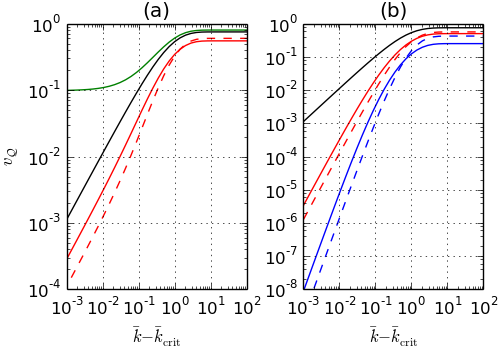}
    \caption{{\bf Simultaneous probabilities for avoiding many colors give a qualitative understanding of critical exponents.} {\bf (a)} With two avoided colors and $r_1=1/4$, $r_2=1/5$, the critical connectivity is $\bar{k}=4/3$. Close to the critical point, $v_{\{1\}}$ grows with exponent of one (black line), saturating to the probability of having color $c\neq1$ for large $\bar{k}$. $v_{\{2\}}$ (solid green line) has a value of about $1/10$ at the critical point. The product of both probabilities $v_{\{1\}} v_{\{2\}}$ (red dashed line) can be seen as an approximation of independent probabilities for $v_{\{1,2\}}\propto S_{\rm color}^{\{1,2\}}$ (red solid line). This helps to qualitatively understand the critical exponent $\beta_k=1$. {\bf (b)} Three avoided colors with $r_1=r_2=r_3=1/4$. Probabilities $v_{\{1\}}$ (black solid line),  $v_{\{1\}}^2$ (red dashed line) and $v_{\{1\}}^3$ (blue dashed line) are useful for a qualitative understanding of the quadratic behavior of $v_{\{1,2\}}$ (red solid line) and cubic behavior of $v_{\{1,2,3\}}\propto S_{\rm color}^{\{1,2,3\}}$ (blue solid line). In this way, the critical exponent $\beta_k=3$ is comprehensible.
    Asymptotic for large $\bar{k}$: $1-r_1-r_2$ true result, $(1-r_1)(1-r_2)=1-r_1-r_2+r_1 r_2$ with independence assumption.}
    \label{fig:v_variables}
\end{center}
\end{figure}

Here  we consider the probability  of a link simultaneously  avoiding all colors in a set $\mathcal{Q}$, or in other words simultaneously connecting to all sets $\mathcal{L}_{\bar{c}}$ for the colors $c\in \mathcal{Q}$.
This probability is calculated exactly in equation \eqref{eq:vu} and is related directly to the overall connectivity (Eq. \eqref{eq:S_color_v}).
When considering the phase transition, we found that this probability is proportional to the product of probabilities  for each color $c\in \mathcal{Q}$ separately.
Although these probabilities are not independent, we found above that the critical point and scaling exponent are consistent with the assumption that they are independent (cf. equation \eqref{eq:v_12_app}).
To understand what impact the dependencies between colors have on the overall color-avoiding connectivity, we compare the assumption that the connections are independent for different colors with the full solution obtained above. 
We define the assumption of independent probabilities (AIP) by taking
\begin{align}
v_{\mathcal{Q}}^{\rm AIP} &\equiv \prod_{q\in\mathcal{Q}} v_{\{q\}}.
\end{align}

As we found for Poisson graphs $S_{\rm color}^{(\mathcal{S},\mathcal{T})} \propto v_{\mathcal{T}}$, this can help us to understand the critical behavior. Apart from that, comparing $v_{\mathcal{Q}}^{\rm AIP}$ with results $v_{\mathcal{Q}}$, where all dependencies are explicitly included, can teach us about dependencies. 

In Fig. \ref{fig:v_variables}a we see that for two avoided colors with different frequencies, $v_{\mathcal{T}}$ and $v_{\mathcal{T}}^{\rm AIP}$ have the same qualitative behavior. This way, the critical exponent $\beta_k$ can be understood as follows: The probability $v_{\{1\}}$ for avoiding a first color has a linear onset starting from the critical point. As the probability $v_{\{2\}}$ already has reached a positive value, the probability for both at the same time has a linear onset as well. With Eq. \ref{eq:v_12_app}, we can estimate dependencies close to the critical point (as long as $v_{\{2\}}\ll1$) to be 
\begin{align}
\frac{v_{\{1,2\}}}{v_{\{1\}} v_{\{2\}}} &\approx \frac{1-r_1-r_2}{(1-r_1) r_2}.
\end{align}
We find that the conditional probability that the same link helps to avoid a second color can be increased as compared to the probability for the first color (as in the figure close to the critical point). The conditional probability can also be suppressed (in the figure for large $\bar{k}$, or for $r_1=0.45$ and $r_2=0.4$ in the whole regime of $\bar{k}$, results not shown). 
In Fig. \ref{fig:v_variables}b we see how $v_{\mathcal{Q}}^{\rm AIP}$ can help to understand larger critical exponents for degenerated largest color frequencies. 

If all colors are avoided, we always have $v_{\{1,\dots,C\}}=0<v_{\{1,\dots,C\}}^{\rm AIP}$. This reflects the fact that in this case no single link can be used to avoid all colors, as the node reached over the link in question has one color, which cannot be avoided. 
Accordingly, only nodes in the two-core can be in the giant CAC component \cite{krause2016hidden}. 
This is in line with our analysis of the limit where $C\rightarrow\infty (N)$~\cite{krause2016hidden}. 
There we showed that, rather than recovering standard percolation in this case, we instead recover $k$-core percolation with $k=2$ \cite{dorogovtsev-prl2006,goltsev-pre2006,lee-preprint2016}.
On the other hand, color connectivity can be expressed in terms of variables $v_{\mathcal{Q}}$ using Eq. \ref{eq:S_color} and \ref{eq:uv}. 
In Sec. \ref{sec:approximations} we discuss that the condition $v_{\{1,\dots,C\}}=0$ must be fulfilled by any appropriate approximation for color avoiding percolation, and how this helps for the heuristic approximation of \cite{krause2016hidden}.

\section{Sender and receiver trust their colors} \label{sec:sender_receiver}

\begin{figure}[htb]
\begin{center}
    \includegraphics[width=1.0\columnwidth]{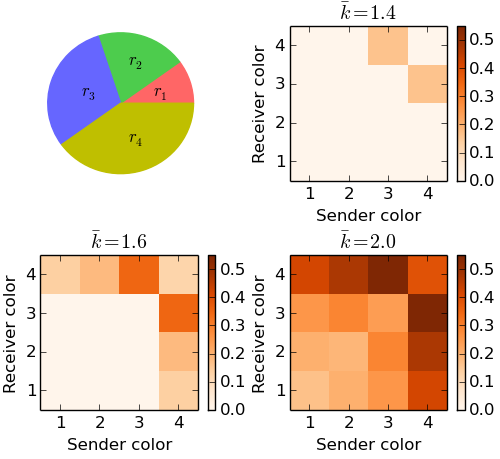}
    \caption{{\bf Sender and receiver trust their own colors.} Having four colors with frequencies $r_c=c/10$, we choose pairs $(c_1,c_2)\subset\{1,2,3,4\}$ of trusted colors for sending and receiving. Shown is the fraction of sender nodes with color $c_1$ which are able to reach nodes of the receiver color $c_2$. With denoting $c_3$ and $c_4$ for the two remaining colors, it reads $S_{\rm color}^{\{c_3,c_4\}}$ (for sender and receiver color being the same $c_1$, we use $c_2$, $c_3$ and $c_4$ denoting the other colors, and $S_{\rm color}^{\{c_2,c_3,c_4\}}$). For the smallest $\bar{k}=1.4$, we see that all nodes of colors one and two are excluded from communication, but they are needed for colors three and four to connect them. For increased $\bar{k}=1.6$, nodes of color four can communicate with nodes of all colors, and with $\bar{k}=2.0$, nodes of all color pairs can communicate.}
    \label{fig:sender_trusted_C4}
\end{center}
\end{figure}
The theory of color-avoiding percolation has been developed to answer the question, what is the maximal set of nodes that can mutually communicate using color-avoiding paths?
Until now, we have examined this question globally, without considering the colors of the sender or receiver nodes.
This is a reasonable assumption for many scenarios, but there are cases where the answer to this central question will depend on the color of the sender and receiver nodes.
For example, if a person in country A wants to communicate with a person in country B, it may be that they both trust the routers of their respective countries, but want to avoid all other countries \cite{krause2016hidden}.  
In such a case, the CAC giant component varies depending on the sender color and receiving color.
This gives rise to a new concept of inter-color connectivity and an inter-color adjacency matrix, as shown in Fig. \ref{fig:sender_trusted_C4}.
We have four colors $c=1,2,3,4$ with different frequencies $r_c=c/10$. This situation is illustrated in the upper left of the figure. On the upper right, we see the fraction among the sender nodes of a color specified on the x-axis, which can reach a macroscopic part of receiver nodes of another color, specified along the y-axis. Results are for Poisson graphs with small $\bar{k}=1.4$. For different trusted colors of sender and receiver nodes, this is $S_{\rm color}^{\{c_3,c_4\}}$ with $c_3, c_4$ being the two colors which are not present on sender and receiver nodes. For example, lets discuss whether nodes of color three can communicate to nodes of color four. The according fraction of nodes having color three is $S_{\rm color}^{\{1,2\}}$, and this is identical to the fraction among nodes of color four which can communicate to color three. We see, that nodes of these both colors are the only ones being CAC. If for example nodes of color two would like to connect to nodes of color four, they need to avoid color three. This is not possible for the small $\bar{k}$ chosen here. With this result we found, that whole classes of nodes (here with colors one and two) can be excluded from CAC, while they are needed for other nodes to provide connectivity. This is in sharp contrast to standard percolation and other variants of percolation questions. It makes sense to allow for transmitting nodes which themselves cannot benefit from CAC. These nodes are connected in the normal giant component and thus functional, only excluded from the more robust color-avoiding connectivity. 

Interestingly, even nodes of color four are not CAC to each other for $\bar{k}=1.4$, as they cannot avoid color three. The according fraction was calculated as $S_{\rm color}^{\{1,2,3\}}$, as trusted color of sender and receiver are both $c=4$, and all other colors have to be avoided. For increased $\bar{k}=1.6$ shown on the lower left of the figure, color three can be avoided, and thus nodes of color four can connect to all other nodes. Still, nodes of other colors are not CAC to each other, as color four cannot be avoided. Finally, with $\bar{k}=2$ (lower right of the figure), nodes of all colors can be CAC, while the fractions for different trusted colors of senders and receivers are still highly heterogeneous. This is different to the case where all colors of a certain set $\mathcal{T}$ always have to be avoided for transmission, even if they are present on the sender and receiver nodes. As a consequence, the fraction of nodes in the largest CAC component is $S^{\mathcal{T}}_{\rm color}$, being the same for all colors. 

\begin{figure}[htb]
\begin{center}
    \includegraphics[width=1.0\columnwidth]{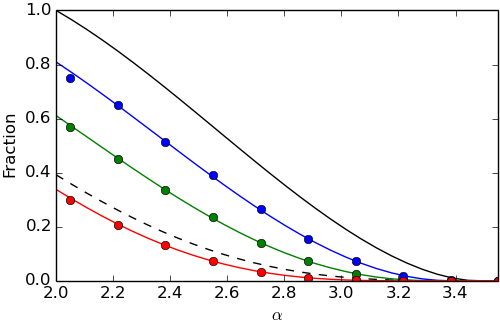}
    \caption{{\bf For scale-free graphs, trusting colors can increase color avoiding connectivity remarkably.} Graphs with broad degree distribution ($p_k\sim k^{-\alpha}$, $k>0$) and the same color frequencies $r_c=c/10$ as in Fig. \ref{fig:sender_trusted_C4}. The red line shows $S_{\rm color}^{\{1,2,3,4\}}$, where no color is trusted for transmission (the circles indicate numerical results, averages over 100 graphs of size $N=10^5$). This is restricted by the two-core (dashed black line). With trusting colors, color-avoidable connectivity is increased above the size of the two-core. Results shown for trusting color four ($S_{\rm color}^{\{1,2,3\}}$, green line and circles) and trusting colors three and four ($S_{\rm color}^{\{1,2\}}$, blue line and circles). The latter case, where sender and receiver nodes of colors three and four trust their colors, is close to full standard connectivity (black line).}
    \label{fig:scale_free}
\end{center}
\end{figure}
Let us finally discuss graphs with broad degree distributions with $p_k\sim k^{-\alpha}$ ($k>0$) and generating functions $g_0(z)={\rm Li}_{\alpha}(z)/\zeta(\alpha)$ and $g_1(z)={\rm Li}_{\alpha-1}(z)/[z\zeta(\alpha-1)]$. ${\rm Li}_{\alpha}(z)$ is the polylogarithm function. In \cite{krause2016hidden} it was pointed out that for such graphs color avoiding connectivity is suppressed when avoiding all colors. This is the case, as the largest CAC component can only be a subset of the two-core. The situation is different, if sender and receiver colors are trusted. This is illustrated in Fig. \ref{fig:scale_free}.

\section{Approximations} \label{sec:approximations}

By discussing the critical behavior for Poisson graphs, we already saw that approximations in calculating $v_{\mathcal{Q}}$ are possible, as long as the defining transcendent equations are developed until sufficient orders. It helps that lower orders cancel out in the defining equations for $v_{\mathcal{Q}}$. Contrarily, for defining equations for $u_{\mathcal{Q}}$, lower order terms do not cancel out, and polynomials of high order have to be solved. While variables $v_{\mathcal{Q}}$ facilitate approximations considerably, there are still some problems left: For degenerated highest frequencies of avoided colors, $v_{\mathcal{Q}}$ has to be approximated up to order $\beta_k$ in order to reflect the critical behavior. For heterogeneous frequencies, the number of sets $\mathcal{Q}$ can be large. Therefore, we discuss two other ways of approximating color avoiding percolation. The first is a heuristic approximation developed in \cite{krause2016hidden}, the second works by redefining avoided colors in order to reduce the number of avoided colors. 

\subsection{Previously published heuristic approximation}

\begin{figure}[htb]
\begin{center}
    \includegraphics[width=1.0\columnwidth]{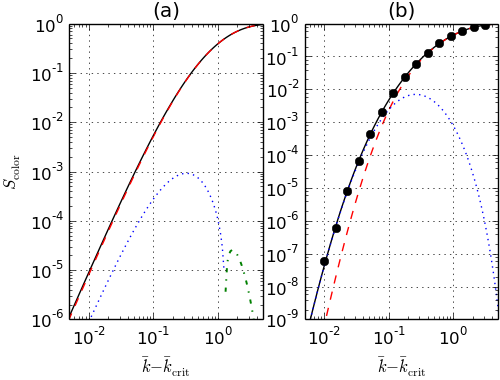}
    \caption{{\bf Comparison with the approximate theory of \cite{krause2016hidden}.} Left: $C=3$ colors of the same frequency $r_c=1/C$ are avoided for transmission, all colors trusted for sending. Results with the approximate theory of \cite{krause2016hidden} (AICP, red dashed line) are close to results with the theory presented here (black solid line). The deviation (positive values with the blue dotted line and negative values shown with the green dash-dotted line) is small. Right: For $C=10$ colors, the deviation is larger, and close to the critical point, the theory of \cite{krause2016hidden} gives poor results, even if the critical behavior is qualitatively captured. Black circles show numerical results (averages over 100 networks of size $N=10^8$).}
    \label{fig:comparison}
\end{center}
\end{figure}

In order to compare with the approximate results presented in \cite{krause2016hidden}, we avoid all $C$ colors for transmission and none for sending. All colors have identical frequencies $r_c=r_1=1/C$. We have $u_{\{c,q\}}=u_{\{1,2\}}$ etc. We can simplify 
\begin{align}
u_{\{1\}} &= r_1 + (1-r_1) g_1(u_{\{1\}}),\\
u_{\{1,2,\dots,j\}} &= j r_1 g_1(u_{\{1,2,\dots,j-1\}}) + (1-j r_1) g_1(u_{\{1,2,\dots,j\}})\nonumber\\
&\quad \quad j=2,\dots,C.\\
S_{\rm color} &= 1 +\sum_{j=1}^{C} (-1)^j {C \choose j} g_0(u_{\{1,2,\dots,j\}}).
\end{align}
This reduces the number of quantities $u_{\mathcal{Q}}$ to be calculated from $2^C-1$ to $C$. This only reduces the computation time, as there are still numerical problems. Combinatorial factors ${C \choose j}$ are large for large $C$, and results for $u_{\{1,2,\dots,j\}}$ can only be numerical, with limited precision. Especially limited precision for calculating the generating functions causes problems for small $S_{\rm color}$. 

In \cite{krause2016hidden}, $u_{\{1,2,\dots,j\}}$ was estimated using the approximation 
\begin{align}
u_{\{1,\dots,j\}} &\approx u+(1-u)\left[\frac{j}{C}(U_{\{1\}})^{j-1} + \frac{C-j}{C}(U_{\{1\}})^{j}\right]\\
&\qquad \equiv u_{\{1,\dots,j\}}^{\rm AICP},\\
U_{\{1\}} &= 1 - \frac{1-u_{\{1\}}}{(1-u)(1-r_c)}.\label{eq:U_c}
\end{align}
$u$ is the probability that a node is not connected to the 
giant component over one particular link and is computed as the solution of $u=g_1(u)$, where $g_1(z)=g_0'(z)/g_0'(1)$
is the generating function of excess degree \cite{newman-book2010}. $U_{\{1\}}$ denotes the conditional probability that a link 
fails to connect to $\mathcal{L}_{\bar 1}$ given that it does connect to the normal giant component via a node having a color $c\neq 1$. We define $U_{\{1\}}=1$ if $u=1$. The probability $u_{\{1\}}$ that a single link does not connect to a giant $\mathcal{L}_{\bar 1}$ 
is calculated with Eq. \ref{Eq:self_cons_1}.

\begin{figure}[htb]
\begin{center}
    \includegraphics[width=1.0\columnwidth]{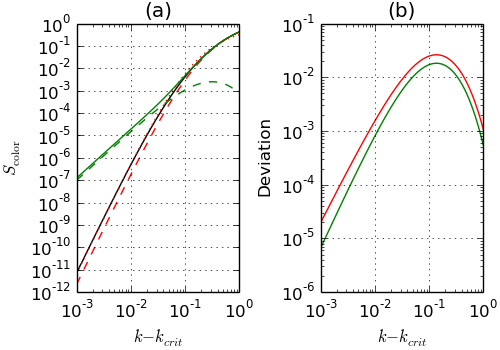}
    \caption{{\bf A fake approximation, closer to the theory then AICP but violating constraint Eq. \ref{eq:constraint}, gives poor results.} Left: For $C=5$ colors, we see results of the theory presented here (black solid line), and results with AICP (red dashed line). Results of a fake approximation are poor (solid green line), because this approximation violates the constraint, residual term $\Delta^{\rm app1}$ shown with a dashed green line. Right: The error $\delta_5^{\rm AICP}$ (red line) made in describing $u_{\{1,\dots,5\}}$ is quadratic in $\bar{k}-\bar{k}_{\rm crit}$ close to the critical point. For the fake approximation, $\delta_5^{\rm app1}$ is constructed to be smaller (green line).}
    \label{fig:approx}
\end{center}
\end{figure}
\begin{figure}[htb]
\begin{center}
    \includegraphics[width=1.0\columnwidth]{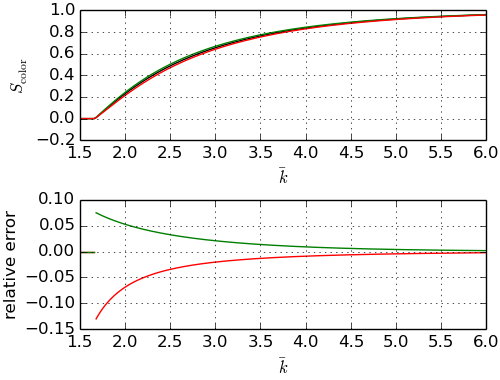}
    \caption{{\bf Uniting colors of small frequency.} The black line shows results, where eight colors are avoided, with frequencies $(r_c)_c=(4/10,2/10,1/10,1/20,1/40,1/80,1/80)$, summing up to 0.8. The green line shows an upper limit, where only the first three colors are avoided, with $(r_c)_c=(4/10,2/10,1/10)$. The red line shows a lower limit, where the last six colors are united into one new color, with $(r_c)_c=(4/10,2/10,2/10)$. Combining many colors of small frequencies into one color, reduces color avoiding connectivity only slightly, thus allowing for better performance of algorithms.}
    \label{fig:approximation}
\end{center}
\end{figure}
The approximation for $u_{\{1,\dots,j\}}$ is motivated as follows: With probability $u$ a link does not connect to the giant component, so $u_{\{1,\dots,j\}}\geq u$. If it connects to the giant component (probability $(1-u)$), then the node on the other side has either one of the $j$ colors (probability $j/C$), leaving simultaneous failure probability of $(U_{\{1\}})^{j-1}$, or it has a different color with probability $(C-j)/C$, leaving simultaneous failure probability of $(U_{\{1\}})^{j}$. In this approximation, we neglect dependencies among the conditional probabilities $U_{\{1\}}, U_{\{2\}}$ etc. Therefore, we call this approximation in the following as the approximation of independent conditional probabilities (AICP). 

A comparison of AICP and the theory presented here is shown in Fig. \ref{fig:comparison}. We see that for $C=3$ colors AICP works well (upper panel). For as many as $C=10$ colors, results are still good for large $S_{\rm color}$, while for $S_{\rm color}<10^{-2}$ there are strong deviations between the theory presented here and AICP. However, the critical behavior in terms of the critical point $\bar{k}_{\rm crit}$, and the fact that the critical exponent $\beta_k$ is large, are still captured. 

In order to understand AICP better, let us define 
\begin{align}
\delta_j^{\rm AICP} &\equiv u_{\{1,\dots,j\}}-u_{\{1,\dots,j\}}^{\rm AICP}.    
\end{align}
In Fig. \ref{fig:approx} on the right, $\delta_5^{\rm AICP}$ is shown with a red solid line, for $C=5$. 
We see that $\delta_5^{\rm AICP}$ is a quadratic function in $\bar{k}-\bar{k}_{\rm crit}$ for small values of this variable, therefore $u_{\{1,\dots,C\}}^{\rm AICP}$ behaves like a linear expansion around the critical point. This is true for $j<C$ as well (results not shown). However, final results for $S_{\rm color}$ as shown on the left of the figure (black line for the theory, red dashed line for AICP) are compatible with an exponent $\beta_k=C=5$. To understand better, how a linear expansion in $\bar{k}-\bar{k}_{\rm crit}$ of variables $u_{\{1,\dots,j\}}$ can finally reproduce such a steep critical behavior, let us discuss the constraints 
\begin{align}
v_{\{1,\dots,C\}}=1 +\sum_{j=1}^{C} (-1)^j {C \choose j} u_{\{1,\dots,j\}} &=0,\label{eq:constraint}\\
1 +\sum_{j=1}^{C} (-1)^j {C \choose j} u_{\{1,\dots,j\}}^{\rm AICP} &=0.
\end{align}
The first of these equations represents the fact that a node can never be CAC via a single link, if no color is trusted. This constraint is also respected with AICP: In \cite{krause2016hidden} it was shown that the largest CAC is always a subset of the 2-core. This constraint also implies that $S_{\rm color}$, as calculated with equation \ref{eq:S_color}, cannot easily be truncated after terms with a certain $|\mathcal{Q}|=Q_{\rm truncate}$. To show that the constraint equation \ref{eq:constraint} is crucial for AICP, let us define a fake approximation denoted as ``app1'', which is closer to the theory, but violates the constraint. We set 
\begin{align}
u_{\{1,\dots,j\}}^{\rm app1}&=u_{\{1,\dots,j\}}-\delta_j^{\rm app1}=u_{\{1,\dots,j\}}-(\delta_j^{\rm AICP})^{1.1}.    
\end{align}
$\delta_5^{\rm app1}$ is shown with a green solid line in the right panel of the figure. As $\delta_j^{\rm AICP}\ll1$ for all $j$, $u_{\{1,\dots,j\}}^{\rm app1}$ is a slightly better approximation of the variables $u_{\{1,\dots,j\}}$. However, the resulting $S_{\rm color}^{\rm app1}$ fails to describe the critical behavior of $S_{\rm color}$, as shown on the left of the figure with a solid green line. Instead we find a result compatible with $S_{\rm color}^{\rm app1}\propto (\bar{k}-\bar{k}_{\rm crit})^2$, what reflects the fact that we have a linear expansion in $\bar{k}-\bar{k}_{\rm crit}$. Indeed, the result is dominated by 
\begin{align}
1 +\sum_{j=1}^{C} (-1)^j {C \choose j} \delta_j^{\rm app1} &\equiv \Delta^{\rm app1}\neq0,
\end{align}
shown on the left of the figure with a dashed green line. If the constraint would be fulfilled, this residual would be zero.

\subsection{Approximation for many small color frequencies}

With the last subsection we saw that the equations defining $S_{\rm color}$ as presented in this paper are hard to approximate. On the other hand, the complexity for homogeneous color frequencies grows exponentially with the number of avoided colors (for transmission). If there is a large number of colors with marginal frequencies, we can use the theory as presented in this paper, and manipulate the set of avoided colors, or combine all colors with small frequencies to one color. As can be seen in Fig. \ref{fig:approximation}, this allows to give upper and lower bounds to $S_{\rm color}$.

\section{Conclusion}

Here we developed a theory for calculating the size of the giant color avoiding connected component, for randomly distributed colors on random network ensembles. We used dependent simultaneous probabilities, that a certain link does or does not enable to avoid several colors at the same time. The conditional probability, that a link helps avoiding a second color $c'$, after it already helps avoiding a first color $c$, can be enhanced or suppressed compared to independent probabilities. An open task for future work would be to understand the mechanics behind these dependencies. Further we found that a clear understanding of simultaneous probabilities helps analyzing the critical behavior. It also helps assessing a previously published heuristic approximation. In general, it is an interesting finding that dependent probabilities for the same link to fulfill different functions can be calculated simultaneously, within the framework of the configuration model. To our knowledge, this is a new direction in percolation theory, with possible applications also beyond color avoiding percolation. 

We developed the theory in a way such that it allows for flexible trust scenarios, where nodes of a certain color can be trusted or avoided for sending/receiving, and for transmission. This allowed us to compare different trust scenarios among each other, and directly with standard percolation. We found that trusting colors for transmission, can remarkably increase color-avoiding connectivity, especially for scale free graphs. A sender node trusting its own color and the color of potential receiver nodes for transmission, can increase its color-avoiding connectivity the most if its own color is dominating in the network. Colors with small frequencies as compared to the dominating color, have a small impact on color-avoiding connectivity. This allowed us to introduce an approximation, where all colors of small frequencies are united into one color. This idea could also be helpful for designing routing algorithms, without keeping track of too many colors.

\begin{acknowledgments}
We acknowledge financial support from the European
Commission FET-Proactive project MULTIPLEX (Grant
No. 317532) and the Italy-Israel NECST project.
M.D. is grateful to the Azrieli Foundation for the award of an Azrieli Fellowship. V.Z. acknowledges support by the H2020 CSA Twinning Project No. 692194, RBI-T-WINNING, and Croatian centers of excellence QuantixLie and Center of Research Excellence for Data Science and Cooperative Systems.
\end{acknowledgments}

\section*{Appendix}

\begin{lstlisting}[language=Python,frame=false,showstringspaces=false,basicstyle=\footnotesize,breakatwhitespace=True,breaklines=true,numbers=right,numbersep=2pt]
"""This Python script is for calculating color avoiding connectivity, as described in the paper: Color-avoiding percolation, by Sebastian M. Krause, Michael M. Danziger, and Vinko Zlatic. See equations (9) and (10). The script contains one exapmle, how the function S_color can be used."""

from pylab import *
import scipy.optimize as opt
      
def subsets_list(a_set):
    """Returns all possible subsets of a_set as list of tuples (including a_set itself, excluding empty set). Ordered by number of elements"""
    N=len(a_set)
    slist=zeros((2**N,N),dtype=int)
    for i in range(1,N+1):
        slist[:,i-1] = 1*((arange(2**N)%(2**i)) > (2**(i-1)-1))
    re_sort=argsort(sum(slist, axis=1))
    slist=slist[re_sort,:]
    slist_=[]
    for q in range(1,len(slist[:,0])):
        Q=tuple(a_set[i] for i in range(len(slist[0,:])) if slist[q,i])
        slist_.append(Q)
    return slist_

def gen_0(z,k):
    """Generating function of degree for Poisson graph."""
    return exp(-k*(1.-z))

def gen_1(z,k):
    """Generating function of excess degree for Poisson graph."""
    return exp(-k*(1.-z))

def fixpoint_u(Q, u, r, k):
    """Calculates u[Q], given vector of color frequencies r and all values u[R] with R subset of Q (except R=Q). k is the average connectivity. See Eq. (9)."""
    sum_r=0.
    sum_r_times_g=0.
    for q in range(len(Q)):
        sum_r+=r[Q[q]]
        Q_=tuple(q2 for q2 in Q if q2!=Q[q])
        sum_r_times_g+=r[Q[q]]*gen_1(u[Q_],k)
    g=lambda x,sum_r,sum_r_times_g,k: sum_r_times_g+(1.-sum_r)*gen_1(x,k)
    u_ = opt.fixed_point(g, 0.5, args=[sum_r,sum_r_times_g,k])
    return u_

def S_color(r,A,k):
    """Calculate S_color, given color frequencies r, first A colors avoided. k is the average connectivity. See Eq. (10)."""
    u={():1}
    S_color_=1.
    for Q in subsets_list(arange(A)):
        u[Q]=1.
        u[Q]=fixpoint_u(Q, u, r, k)
        S_color_+=(-1.)**len(Q)*gen_0(u[Q],k)
    return S_color_

### Calculate S_color for a special case, compare Fig. 2(a).
### Result should be about 0.5054053680
r=array([0.5,0.3,0.2]) # Color frequencies
A=2 # Avoid first two colors
k=3 # Average degree
result=S_color(r,A,k)
print(result)

\end{lstlisting}

\bibliography{mp}

\end{document}